\documentclass[useAMS,usenatbib]{mn2e}
\usepackage{graphicx}

\def\beq{\begin{equation}}
\def\eeq{\end{equation}}

\def\masyr{I_{\rm mas\,{yr}^{-1}}}

\def\sigmal{\sigma_{\rm l}}
\def\sigmab{\sigma_{\rm b}}
\def\mul{\mu_{\rm l}}
\def\mub{\mu_{\rm b}}

\def\kms{{\,\rm km\,s^{-1}}}
\def\mas{{\rm \ mas}}
\def\masyr{{\rm\mas\ yr}^{-1}}

\newcommand\aj{{AJ}}%
\newcommand\apj{{ApJ}}%
\newcommand\apjl{{ApJ}}%
\newcommand\apjs{{ApJS}}%
\newcommand\mnras{{MNRAS}}%
\newcommand\pasp{{PASP}}%
\date{Accepted 2006 April 26. Received 2006 April 26; in original form 2006 March 22}

\pubyear{2006}

\title[]{Mapping stellar kinematics across the Galactic bar:\\
HST measurements of proper motions in 35 fields.
\thanks{
Based on observations made with the NASA/ESA Hubble Space Telescope,
obtained from the data archive at the Space Telescope Science Institute. 
STScI is operated by the Association of Universities for Research in Astronomy, 
Inc. under NASA contract NAS 5-26555.}}

\author[Koz{\l}owski et al.]{\newauthor
S. Koz{\l}owski$^1$\thanks{E-mail: \tt simkoz@jb.man.ac.uk}, P. R. Wo\'zniak$^2$, S. Mao$^1$,
M. C. Smith$^3$, T. Sumi$^4$,\newauthor
W. T. Vestrand$^2$ and {\L.} Wyrzykowski$^5$\\
\\
$^1$ Jodrell Bank Observatory, The University of Manchester, Macclesfield, Cheshire SK11 9DL, UK\\
$^2$ Los Alamos National Laboratory, MS-D466, Los Alamos, NM 87545, USA\\
$^3$ Kapteyn Institute, PO Box 800, 9700 AV Groningen, the Netherlands\\
$^4$ Princeton University Observatory, Princeton, NJ 08544, USA\\
$^5$ Institute of Astronomy, University of Cambridge, Madingley Road, Cambridge CB3 OHA, UK}

\begin{document}

\maketitle

\begin{abstract}
We present a proper motion mini-survey of 35 fields in the vicinity of Baade window, $(l,b) = (1^\circ,-4^\circ)$, sampling roughly
a $5\times2.5$ deg region of the Galactic bar. Our second epoch observations collected with the ACS/HRC instrument on board
the Hubble Space Telescope were combined with the archival WFPC2/PC images.
The resulting time baselines are in the range of 4--8 years. Precise proper motions of 15,863 stars were determined in the 
reference frame defined by the mean motion of stars with magnitudes between $I_{F814W} = 16.5 - 21.5$ along the line of sight.
We clearly detect small gradients in proper motion dispersions $(\sigmal,\sigmab) \sim (3.0, 2.5) \masyr$, and in the amount
of anisotropy ($\sigmal/\sigmab \sim 1.2$). Both the longitude dispersion $\sigmal$ and its ratio to the vertical motion
$\sigmab$ increase toward the Galactic plane. The decline of the anisotropy ratio $\sigmal/\sigmab$ toward the minor axis
of the bulge is mostly due to increasing $\sigmab$. We also find, for the first time, a significant negative covariance
term in the transverse velocity field $\sigma_{\rm lb}/(\sigmal\sigmab) \simeq -0.10$. Our results extend by a factor of
$\sim$15 the number of the Galactic bar fields with good proper motion dispersions.
\end{abstract}


\section{Introduction}
\label{sec:intro}

The Milky Way appears to be a typical spiral galaxy with a disk and bulge. While our unique inside
view of the Galaxy helps to understand the galactic structure in general, it also makes it more difficult 
to identify structures such as bars. The case for existence of a bar at the Galactic centre -- first proposed 
by \cite{deV64} -- is easier to make knowing that bars are common in external galaxies.

There is now conclusive evidence that the Galactic bulge is of a barred type. The longitude asymmetry
of the COBE photometric maps (\citealt{BS91,Dwe95}), high optical depths to gravitational microlensing
(\citealt{ZSR95}), asymmetric star counts (\citealt{Sta94,BG05}), non-circular gas kinematics (\citealt{deV64}),
and triaxiality of the stellar velocity field (\citealt{ZSR94,ZRB96}) have all been interpreted as signatures
of the Galactic bar. Unfortunately, the size and precise orientation of the bar are still being debated.
Recently \cite{Ben05} found that the infra-red star counts collected by the Spitzer Space Telescope are best
explained assuming a bar with a half-length $4.4\pm0.5$ kpc placed at a $\sim 44^\circ$ angle to the Sun--Galactic
center line. Most previous studies prefer a bar at $\sim 20^\circ$ extending out to $\sim$3.5 kpc (e.g. \citealt{Ger01}).
Such apparently conflicting evidence may be an indication that the inner Galaxy hides even more complicated
structures. A secondary bar (\citealt{Ala01,BG05}) and a ring (\citealt{SK01}) have been suggested, since these
features are also evident in many other spiral galaxies (e.g. \citealt{ES99}).

\cite{Bin05} discussed the progress of the dynamical modeling techniques in the context of major observational
advances expected from a future space mission GAIA. The three approaches to constructing a self-consistent
dynamical Galaxy model are the Schwarzschild method (\citealt{ZSR94,Haf00}), 
the torus modelling method (see \citealt{Bin05} for details) and
$N$-body simulations with particle weights determined by the ``made-to-measure'' algorithm
(\citealt{ST96}).
The first Galactic bar model employing the latter method was built by \cite{Bis04}. Neither of these techniques can fully address
the structure of the inner Galaxy without constraints on stellar kinematics. The refinement of the models
is limited largely by the scarcity of good proper motion and radial velocity measurements. \cite{Bis04}, for example,
compared kinematic predictions of their model with the data for just two lines of sight. A handful of samples
published since the pioneering photographic work of \cite{Spa92} is not enough to remove the non-uniqueness
of the model parameters.

In a study based on two lines of sight \cite{KR02} have demonstrated that high quality relative proper motions
can be obtained with a relatively modest investment of time using the Hubble Space Telescope (HST). At the resolution
of the Wide Field Planetary Camera 2 (WFPC2) instrument the required time baseline is only a few years. The HST archive
contains a number of images suitable as the first epoch data, so the tedious part of accumulating the baseline
can be avoided entirely. Most of these fields are centred around microlensing events discovered
by the MACHO collaboration (e.g. \citealt{Pop05}). Using a similar concept to that of \cite{KR02}, we carried out a mini-survey
of proper motions in 35 of the available MACHO fields to study the kinematics of microlensed sources and of the 
general stellar populations. Here we present the measurement techniques and results for the general stellar 
population in these 35 lines of sight.

\begin{table*}
\begin{centering}
\begin{minipage}{\textwidth}
\caption{Log of the HST observations.}
\vspace{0.2cm}
\begin{tabular}{lcccllcccl}
\hline
& & & \multicolumn{4}{c}{First epoch} & & \multicolumn{2}{c}{Second epoch} \\*[0.1cm]
\cline{4-7} \cline{9-10}\\*[-0.2cm]
\makebox[2.1cm][l]{MACHO field} & RA & Dec & Year & F814W exp. & F555W exp. & Prop. ID & & Year & F814W exp. \\
\hline 
108-C      \dotfill  & 18:00:01.276 & $-$28:27:41.23 & 1996.82 & 6 $\times$ 260 s &                  & 6756 & & 2005.16 & 4 $\times$ 260 s \\ 
119-C      \dotfill  & 18:03:03.010 & $-$30:09:56.50 & 1996.82 & 6 $\times$ 260 s &                  & 6756 & & 2005.15 & 4 $\times$ 260 s \\ 
119-D 	   \dotfill  & 18:04:24.825 & $-$30:05:58.94 & 1996.82 & 6 $\times$ 260 s &                  & 6756 & & 2004.78 & 4 $\times$ 260 s \\ 
120-A 	   \dotfill  & 18:07:26.441 & $-$29:39:34.22 & 1996.82 & 6 $\times$ 260 s &                  & 6756 & & 2005.15 & 4 $\times$ 260 s \\ 
167-A  	   \dotfill  & 18:13:32.154 & $-$26:31:10.33 & 1996.82 & 6 $\times$ 260 s &                  & 6756 & & 2005.16 & 4 $\times$ 260 s \\ 
101-C 	   \dotfill  & 18:07:32.649 & $-$27:31:35.60 & 1997.47 & 6 $\times$ 260 s &                  & 6756 & & 2005.15 & 4 $\times$ 260 s \\ 
95-BLG-11  \dotfill  & 18:04:37.239 & $-$30:12:11.45 & 1996.67 & 6 $\times$ 260 s &                  & 6756 & & 2005.16 & 4 $\times$ 260 s \\ 
96-BLG-17  \dotfill  & 18:06:09.107 & $-$27:53:38.59 & 1996.81 & 6 $\times$ 260 s &                  & 6756 & & 2005.17 & 4 $\times$ 260 s \\ 
119-A 	   \dotfill  & 18:03:35.789 & $-$29:42:01.26 & 1996.68 & 6 $\times$ 160 s & 2 $\times$ 400 s & 6756 & & 2005.14 & 4 $\times$ 160 s \\ 
95-BLG-7   \dotfill  & 18:13:29.298 & $-$26:13:58.12 & 1998.84 & 2 $\times$ 40 s  & 3 $\times$ 40 s  & 7431 & & 2005.46 & 4 $\times$ 40 s  \\ 
95-BLG-10  \dotfill  & 17:58:16.011 & $-$29:32:10.86 & 1997.82 & 2 $\times$ 40 s  & 3 $\times$ 40 s  & 7431 & & 2004.66 & 4 $\times$ 40 s  \\
95-BLG-13  \dotfill  & 18:08:47.038 & $-$27:40:47.25 & 1999.45 & 2 $\times$ 40 s  & 3 $\times$ 40 s  & 7431 & & 2005.12 & 4 $\times$ 40 s  \\
95-BLG-14  \dotfill  & 18:01:26.308 & $-$28:31:14.03 & 2000.45 & 2 $\times$ 40 s  & 3 $\times$ 40 s  & 7431 & & 2005.40 & 4 $\times$ 40 s  \\
95-BLG-19  \dotfill  & 18:11:32.487 & $-$27:45:26.99 & 1998.49 & 2 $\times$ 40 s  & 3 $\times$ 40 s  & 7431 & & 2005.44 & 4 $\times$ 40 s  \\
97-BLG-18  \dotfill  & 18:03:15.254 & $-$28:00:14.06 & 1998.59 & 2 $\times$ 40 s  & 3 $\times$ 40 s  & 7431 & & 2005.31 & 4 $\times$ 40 s  \\
104-C 	   \dotfill  & 18:03:34.050 & $-$28:00:18.94 & 1998.73 & 2 $\times$ 40 s  & 3 $\times$ 40 s  & 7431 & & 2005.43 & 4 $\times$ 40 s  \\
104-D 	   \dotfill  & 18:03:29.024 & $-$28:00:30.99 & 1998.80 & 2 $\times$ 40 s  & 3 $\times$ 40 s  & 7431 & & 2005.45 & 4 $\times$ 40 s  \\
108-A 	   \dotfill  & 18:00:25.866 & $-$28:02:35.24 & 1998.80 & 2 $\times$ 40 s  & 3 $\times$ 40 s  & 7431 & & 2005.16 & 4 $\times$ 40 s  \\
128-B 	   \dotfill  & 18:07:18.624 & $-$28:59:29.83 & 1998.49 & 2 $\times$ 30 s  & 3 $\times$ 40 s  & 7431 & & 2005.37 & 4 $\times$ 30 s  \\
104-B 	   \dotfill  & 18:03:09.046 & $-$28:01:45.25 & 1999.45 & 2 $\times$ 40 s  & 3 $\times$ 40 s  & 7431 & & 2005.39 & 4 $\times$ 40 s  \\
128-A 	   \dotfill  & 18:06:57.621 & $-$29:00:55.15 & 1999.33 & 2 $\times$ 40 s  & 3 $\times$ 40 s  & 7431 & & 2005.49 & 4 $\times$ 40 s  \\
94-BLG-3   \dotfill  & 17:58:25.300 & $-$29:47:59.50 & 1997.82 & 2 $\times$ 40 s  & 3 $\times$ 40 s  & 7431 & & 2005.48 & 4 $\times$ 40 s  \\ 
94-BLG-4   \dotfill  & 17:58:36.766 & $-$30:02:19.27 & 1997.82 & 2 $\times$ 40 s  & 3 $\times$ 40 s  & 7431 & & 2005.16 & 4 $\times$ 40 s  \\ 
95-BLG-36  \dotfill  & 18:07:20.775 & $-$27:24:09.69 & 1998.80 & 2 $\times$ 40 s  & 3 $\times$ 40 s  & 7431 & & 2005.39 & 4 $\times$ 40 s  \\
95-BLG-37  \dotfill  & 18:04:34.452 & $-$28:25:33.46 & 1999.43 & 2 $\times$ 40 s  & 3 $\times$ 40 s  & 7431 & & 2004.67 & 4 $\times$ 40 s  \\
95-BLG-38  \dotfill  & 17:59:41.851 & $-$28:12:10.31 & 1998.81 & 2 $\times$ 40 s  & 3 $\times$ 40 s  & 7431 & & 2005.33 & 4 $\times$ 40 s  \\
95-BLG-41  \dotfill  & 18:02:06.332 & $-$28:50:45.26 & 1999.46 & 2 $\times$ 40 s  & 3 $\times$ 40 s  & 7431 & & 2005.44 & 4 $\times$ 40 s  \\
96-BLG-14  \dotfill  & 18:05:15.421 & $-$27:58:25.01 & 1997.83 & 2 $\times$ 40 s  & 3 $\times$ 40 s  & 7431 & & 2004.67 & 4 $\times$ 40 s  \\
96-BLG-4   \dotfill  & 18:06:11.954 & $-$28:16:52.77 & 1998.79 & 2 $\times$ 26 s  & 3 $\times$ 40 s  & 7431 & & 2004.82 & 4 $\times$ 26 s  \\
97-BLG-38  \dotfill  & 18:04:06.083 & $-$27:48:26.25 & 1998.51 & 2 $\times$ 40 s  & 3 $\times$ 40 s  & 7431 & & 2004.63 & 4 $\times$ 40 s  \\
97-BLG-24  \dotfill  & 18:04:20.253 & $-$27:24:45.28 & 1998.35 & 2 $\times$ 40 s  & 3 $\times$ 40 s  & 7431 & & 2005.49 & 4 $\times$ 40 s  \\
96-BLG-5   \dotfill  & 18:05:02.497 & $-$27:42:17.23 & 1999.45 & 4 $\times$ 160 s & 2 $\times$ 400 s & 8490 & & 2005.12 & 4 $\times$ 160 s \\ 
98-BLG-6   \dotfill  & 17:57:32.812 & $-$28:42:45.41 & 2000.48 & 2 $\times$ 100 s & 2 $\times$ 260 s & 8654 & & 2004.73 & 4 $\times$ 100 s \\ 
97-BLG-41  \dotfill  & 17:56:20.691 & $-$28:47:41.97 & 2000.47 & 4 $\times$ 100 s & 4 $\times$ 160 s & 8654 & & 2004.62 & 4 $\times$ 100 s \\
99-BLG-22  \dotfill  & 18:05:05.281 & $-$28:34:41.69 & 2001.77 & 4 $\times$ 400 s & 4 $\times$ 400 s & 9307 & & 2005.16 & 4 $\times$ 400 s \\

\hline
\end{tabular}
\end{minipage}
\end{centering}
\label{tab:data}
\end{table*}

\section{HST images and data reduction}
\label{sec:data}

The log of the HST observations used in our proper motion mini-survey is given in Table~\ref{tab:data}.
The first epoch images (selected from the HST archive\footnote{\tt http://archive.stsci.edu/hst/})
were all taken with the WFPC2/PC camera, and cover the time interval 1996--2000. The second epoch data
come from our own SNAP program (cycle 13; proposal ID 10198) and were collected in 2004 and 2005
using the High Resolution Channel (HRC) of the Advanced Camera for Surveys (ACS). Our SNAP survey was optimized
toward high execution rates and, therefore, we only requested F814W
observations to keep the required target visibility
as low as possible. Both PC and HRC detectors cover a similar field of view ($25\arcsec\times29\arcsec$ and $35\arcsec\times35\arcsec$,
respectively) and have comparable pixel scales ($27$ versus $45.5$ mas). There were no restrictions
on the telescope roll angle during ACS observing. While the latter relaxed condition decreased the number of possible
proper motion determinations, it greatly improved scheduling opportunities.
Most of the subsequent analysis for each of the 35 fields is based on a pair of good quality
F814W ($I$-band) images constructed by stacking all suitable data for a given epoch. In some cases,
the first epoch data included F555W ($V$-band) images that allowed us to construct color-magnitude diagrams (CMDs).
We also re-analyzed the two fields previously studied by \cite{KR02}, increasing to 37 the total number
of the Galactic bulge fields considered here.

\subsection{Image reductions}
\label{sec:image}

The basic reductions of the ACS images, i.e. de-biasing, dark frame subtraction, flat-fielding, and cosmic-ray
removal, were performed on-the-fly by the standard HST data processing pipeline. The pipeline also takes care
of dithering, cosmic-ray splits and geometric corrections using the Multidrizzle software (\citealt{Koe02}), 
which in turn uses the Drizzle routines (\citealt{FH02}). Our ACS observations employed a generous
4-point dithering pattern combined with a 2-way cosmic-ray split, providing the final drizzled images with high S/N ratio,
excellent dynamic range and highly reliable cosmic-ray rejection. In case of the first-epoch WFPC2 images we utilized
the standard HST data products for individual exposures, and then used the {\tt drizzle} task of the IRAF package
to correct the geometric distortions. For cosmic-ray cleaning, registering and combining these corrected images
we developed dedicated IRAF scripts. The quality of our final cross-instrument astrometry is limited
by the larger pixel size, as well as the lower S/N ratio and number of the individual first epoch PC frames
available for stacking by comparison to the ACS data (see \S\,\ref{sec:pm}).

\subsection{Object catalogs and PSF fitting}
\label{sec:psf}

The instrumental positions and magnitudes of the field objects were measured using the IRAF task {\tt starfind},
an improved version of {\tt daofind} that fits Gaussian profiles to stellar images. The combined images from both
WFPC2/PC and ACS/HRC detectors have a well sampled point-spread-function (PSF) with the full-width-at-half-maximum (FWHM) of stellar images,
correspondingly, 2.4 and 2.8 pixels. Our PSF fits were restricted within the area of the Airy disc (3.0 and 2.0 pixel
radius for PC and HRC data respectively), where the point source flux is well approximated by a Gaussian model.
Outside the Airy disc the PSFs show a variety of shapes, including rings, possible diffraction spikes and bright
spots in case of high S/N objects. These features can mimic stars and need to be carefully considered during object
cross-identification. We imposed a minimum separation of 3$\times$FWHM between any two sources detected in the same
image to ensure that there are essentially no spurious objects in the final source lists. The loss of number statistics
due to the accidental rejection of the actual stars in tight groups is insignificant.
In fact, the centroid measurements for objects in the wings of other stars are notoriously unreliable and best avoided.
The minimum separation cut also helps in cross identification of objects between the two epochs (\S\,\ref{sec:pm}),
since the expected intrinsic object shifts may reach $\sim$2 PC pixels.

The final object catalogs were converted to the VEGA magnitude system (\citealt{Gir02} and references therein) and the 
astrometric transformations to
the Galactic (l, b) coordinates were established using the World Coordinate System (WCS) headers of the ACS images.
Our estimated S/N ratios for object fluxes are based on propagated errors in pixel counts that account for photon statistics.

\section{Estimating transverse motions of the Galactic bulge stars}
\label{sec:pm}

Absolute astrometry is difficult in the crowded Galactic bulge fields. Until we can establish a sample of extragalactic objects
(e.g. spectroscopically confirmed QSOs in the catalogue of candidates by \citealt{Sum05})
shining through the low extinction windows, the only readily available reference velocity
in the Galactic bulge is the mean velocity of stars along the line of sight. 
Notice, however, that the second order moments of proper motions are unaffected by the choice of our reference frame.
In this analysis we use the magnitude- and distance-selected samples to investigate the spatial dependence of the covariance
matrix of the transverse velocity field across the Galactic bar.

\subsection{Relative proper motions and their dispersions}
\label{sec:pm_and_disp}

Having measured the instrumental positions of stars on both the first and the second epoch images, we tied the
WFPC2/PC positions to the ACS/HRC pixel grid. The object shifts $\Delta \rm l$ and $\Delta \rm b$ in the Galactic coordinates
between the two epochs could then be calculated using the WCS information from the ACS headers. 
We cross-correlated the positions of a few hundred stars in the magnitude range $16.5 < I_{\rm F814W} < 21.5$
to obtain the coordinate transformation between two pixel grids, which is approximated by a third order polynomial.
Stars brighter than $I_{\rm F814W}=16.5$ were often
saturated while those with $I_{\rm F814W} > 21.5$ were too faint to have useful S/N ratios, particularly for the fields
with short exposures (Table~\ref{tab:data}). Our procedure for cross-identifying stars
starts from matching the first 20 objects (out of $\sim$50 brightest stars with $17 < I_{\rm F814W} < 18$) using the triangle
algorithm (\citealt{Gro86}; \citealt{W00}). The initial low-order fit is then iteratively refined. A star with a transverse
velocity of 100 km/s at the distance of 8 kpc will move by 26.4 mas, or roughly one ACS/HRC pixel, assuming a 10 year baseline.
Accordingly, we adopted a tolerance radius of 100 mas for the final matching.

After geometrically aligning and transforming object positions to the Galactic (l, b) coordinates, we folded the data
with the time baseline and estimated all components of the transverse velocity tensor, i.e. dispersions $\sigmal$, $\sigmab$
and the normalized covariance $C_{\rm lb} \equiv \sigma_{\rm lb}/(\sigmal\sigmab)$. The sample of stars used to trace the kinematic
parameters of the Galactic bulge was limited to the magnitude range $18.0 < I_{\rm F814W} < 21.5$, i.e. dominated by the bulge main sequence
population near the turn-off point. This puts all lines of sight (with data sets of the varying depth and dynamic range)
on a more common footing. However, as already noted by \cite{KR02}, the results are insensitive to the details
of the magnitude cuts.

\subsection{Astrometric errors}
\label{sec:errors}

The 1$\sigma$ centroid errors from PSF fitting (per coordinate)
can be estimated from the S/N ratio (SNR):

\begin{equation}
\label{eq:errors}
\delta \simeq \gamma \times {{\rm FWHM} \over {\rm SNR}},
\end{equation}

\noindent
where $\gamma = 0.6$ for a Gaussian PSF model and the FWHM is in pixels (see e.g. \citealt{KR02}). We tested this prescription by stacking
independent subsets of images taken at a single epoch. Similarly to \cite{KR02}, we find that Equation~\ref{eq:errors} is an excellent
representation of the actual astrometric uncertainties in our data, with the exception of the brightest stars, for which a constant
systematic contribution of 0.025 pixel is required. Consequently, we used Equation~\ref{eq:errors} with the systematic term added
in quadrature to estimate the astrometric errors and their contribution to the apparent proper motion dispersions. The formulas for
estimating $\sigmal$, $\sigmab$ and their errors corrected for the measurement variance can be found in \cite{Spa92}.
Throughout this paper we use bootstrapped uncertainties of the sample statistics (from 1000 trials) that turned out
to be slightly more conservative than analytical formulas. The estimated intrinsic
dispersions reported in \S\,\ref{sec:results} are 5--10\% lower compared to the raw values. The cross term $C_{\rm lb}$ need not be corrected,
as long as the errors in l and b are uncorrelated. None of our conclusions depend on the precise value or even the presence of this correction.

The limiting S/N ratio for a useful detection in our analysis is about 10 and corresponds to a $I_{\rm F814W} \simeq 21.5$ mag star in the combined image of two
40-second WFPC2/PC exposures. The same star will be detected at S/N $\sim$ 20 in the lowest quality ACS stack
(four 40-second frames). The shortest time baseline in our data is 3.388 years, and the typical 1$\sigma$ astrometric
uncertainties for a 21.5 mag star are $\sim$7.2 and $\sim$2.1 mas in the first and the second epoch images respectively. In this worst case scenario,
the proper motion can be measured to an accuracy 2.5 $\masyr$. The images for the first eight fields in Table~\ref{tab:data} have
relatively long exposure times, so the resulting proper motion errors are only $\sim$0.1 $\masyr$ for bright stars
and $\sim$0.3 $\masyr$ for the faintest stars in those samples, with a
systematic error of 0.025 pixel (c.f. discussion following equation~(\ref{eq:errors})).

\begin{table*}
\begin{centering}
\begin{minipage}{13cm}
\caption{\label{tab:pm_results}Results of our proper motion mini-survey. The dispersions $\sigmal, \sigmab$ and the dimensionless
correlation coefficient $C_{\rm lb}$ were measured for 35 lines of sight in the Galactic bulge (l, b).
The time baseline $\Delta t$ and the number of stars $N_{\rm stars}$ used to estimate the kinematics are also given.}
\begin{tabular}{lccccccc}\hline
\makebox[2.1cm][l]{Field name} & l & b & $\sigmal$ & $\sigmab$ & $C_{\rm lb}$ & $\Delta t$ & $N_{\rm stars}$ \\
 & [deg] & [deg] & [$\masyr$] & [$\masyr$] & & [yr] & \\ \hline
      101-C \dotfill &   3.65 & $-$3.47 & $2.85 \pm 0.09$ & $2.45 \pm 0.08$ & $-0.15 \pm 0.05$ &    7.683 &   445 \\
      104-B \dotfill &   2.73 & $-$2.87 & $2.97 \pm 0.10$ & $2.50 \pm 0.10$ & $-0.05 \pm 0.05$ &    5.941 &   407 \\
      104-C \dotfill &   2.80 & $-$2.93 & $2.74 \pm 0.09$ & $2.51 \pm 0.10$ & $-0.15 \pm 0.04$ &    6.706 &   482 \\
      104-D \dotfill &   2.79 & $-$2.92 & $2.84 \pm 0.10$ & $2.36 \pm 0.10$ & $-0.10 \pm 0.05$ &    6.649 &   437 \\
      108-A \dotfill &   2.42 & $-$2.35 & $2.90 \pm 0.12$ & $2.32 \pm 0.12$ & $-0.08 \pm 0.06$ &    6.360 &   396 \\
      108-C \dotfill &   2.02 & $-$2.48 & $3.15 \pm 0.10$ & $2.52 \pm 0.07$ & $-0.09 \pm 0.04$ &    8.345 &   615 \\
      119-A \dotfill &   1.32 & $-$3.77 & $2.89 \pm 0.10$ & $2.44 \pm 0.08$ & $-0.14 \pm 0.04$ &    8.458 &   471 \\
      119-C \dotfill &   0.85 & $-$3.89 & $2.79 \pm 0.10$ & $2.65 \pm 0.08$ & $-0.14 \pm 0.04$ &    8.339 &   459 \\
      119-D \dotfill &   1.06 & $-$4.12 & $2.75 \pm 0.10$ & $2.56 \pm 0.09$ & $-0.05 \pm 0.06$ &    7.962 &   420 \\
      120-A \dotfill &   1.76 & $-$4.48 & $2.75 \pm 0.09$ & $2.52 \pm 0.09$ & $-0.04 \pm 0.05$ &    8.339 &   397 \\
      128-A \dotfill &   2.28 & $-$4.08 & $2.63 \pm 0.11$ & $2.33 \pm 0.12$ & $-0.12 \pm 0.05$ &    6.165 &   357 \\
      128-B \dotfill &   2.33 & $-$4.13 & $2.70 \pm 0.12$ & $2.29 \pm 0.13$ & $-0.13 \pm 0.06$ &    6.881 &   338 \\
      167-A \dotfill &   5.17 & $-$4.16 & $2.75 \pm 0.11$ & $2.36 \pm 0.09$ & $-0.18 \pm 0.05$ &    8.345 &   317 \\
   94-BLG-3 \dotfill &   0.68 & $-$2.84 & $2.84 \pm 0.10$ & $2.58 \pm 0.10$ & $-0.12 \pm 0.05$ &    7.654 &   496 \\
   94-BLG-4 \dotfill &   0.49 & $-$3.00 & $2.58 \pm 0.11$ & $2.46 \pm 0.09$ & $-0.03 \pm 0.04$ &    7.341 &   413 \\
  95-BLG-10 \dotfill &   0.89 & $-$2.68 & $3.07 \pm 0.10$ & $2.41 \pm 0.09$ & $-0.12 \pm 0.04$ &    6.840 &   487 \\
  95-BLG-11 \dotfill &   0.99 & $-$4.21 & $2.82 \pm 0.09$ & $2.62 \pm 0.09$ & $-0.14 \pm 0.04$ &    8.493 &   443 \\
  95-BLG-13 \dotfill &   3.64 & $-$3.78 & $2.61 \pm 0.13$ & $2.31 \pm 0.12$ & $-0.14 \pm 0.05$ &    5.672 &   309 \\
  95-BLG-14 \dotfill &   2.12 & $-$2.78 & $2.95 \pm 0.13$ & $2.50 \pm 0.11$ & $-0.12 \pm 0.05$ &    4.950 &   463 \\
  95-BLG-19 \dotfill &   3.87 & $-$4.36 & $2.61 \pm 0.11$ & $2.17 \pm 0.10$ & $-0.13 \pm 0.06$ &    6.952 &   300 \\
  95-BLG-36 \dotfill &   3.73 & $-$3.37 & $2.75 \pm 0.12$ & $2.11 \pm 0.11$ & $-0.10 \pm 0.05$ &    6.587 &   376 \\
  95-BLG-37 \dotfill &   2.54 & $-$3.33 & $2.72 \pm 0.12$ & $2.44 \pm 0.12$ & $-0.04 \pm 0.05$ &    5.238 &   442 \\
  95-BLG-38 \dotfill &   2.20 & $-$2.29 & $2.87 \pm 0.12$ & $2.46 \pm 0.10$ & $-0.05 \pm 0.04$ &    6.526 &   474 \\
  95-BLG-41 \dotfill &   1.91 & $-$3.07 & $2.79 \pm 0.10$ & $2.34 \pm 0.10$ & $-0.04 \pm 0.05$ &    5.980 &   450 \\
   95-BLG-7 \dotfill &   5.42 & $-$4.01 & $2.86 \pm 0.14$ & $1.88 \pm 0.11$ & $-0.20 \pm 0.07$ &    6.616 &   265 \\
  96-BLG-14 \dotfill &   3.01 & $-$3.24 & $2.71 \pm 0.12$ & $2.40 \pm 0.12$ & $-0.17 \pm 0.05$ &    6.833 &   373 \\
  96-BLG-17 \dotfill &   3.17 & $-$3.38 & $3.07 \pm 0.10$ & $2.55 \pm 0.09$ & $-0.16 \pm 0.04$ &    8.364 &   557 \\
   96-BLG-4 \dotfill &   2.84 & $-$3.57 & $2.68 \pm 0.14$ & $2.26 \pm 0.14$ & $-0.04 \pm 0.06$ &    6.027 &   329 \\
   96-BLG-5 \dotfill &   3.22 & $-$3.07 & $3.17 \pm 0.10$ & $2.39 \pm 0.08$ & $-0.13 \pm 0.05$ &    5.670 &   535 \\
  97-BLG-18 \dotfill &   2.77 & $-$2.87 & $2.99 \pm 0.10$ & $2.38 \pm 0.10$ & $-0.12 \pm 0.04$ &    6.713 &   433 \\
  97-BLG-24 \dotfill &   3.40 & $-$2.79 & $3.00 \pm 0.11$ & $2.39 \pm 0.10$ & $-0.10 \pm 0.05$ &    7.115 &   398 \\
  97-BLG-38 \dotfill &   3.03 & $-$2.94 & $2.95 \pm 0.12$ & $2.21 \pm 0.10$ & $-0.06 \pm 0.05$ &    6.118 &   395 \\
  97-BLG-41 \dotfill &   1.32 & $-$1.95 & $2.58 \pm 0.07$ & $2.13 \pm 0.07$ & $-0.09 \pm 0.04$ &    5.145 &   612 \\
   98-BLG-6 \dotfill &   1.53 & $-$2.13 & $3.26 \pm 0.10$ & $2.79 \pm 0.12$ & $-0.07 \pm 0.05$ &    4.252 &   670 \\
  99-BLG-22 \dotfill &   2.46 & $-$3.50 & $3.11 \pm 0.10$ & $2.60 \pm 0.09$ & $-0.17 \pm 0.04$ &    3.388 &   493 \\
 & & & & & & & \\
      KR-BW\footnote{KR-BW and KR-SgrI are the Baade window and the Sagittarius-I field from \cite{KR02}}
\dotfill &   1.14 & $-$3.77 & $2.87 \pm 0.08$ & $2.59 \pm 0.08$ & $-0.07 \pm 0.03$ &    6.048 &   694 \\
    KR-SgrI \dotfill &   1.26 & $-$2.66 & $3.07 \pm 0.08$ & $2.73 \pm 0.07$ & $-0.09 \pm 0.04$ &    5.960 &   752 \\
\hline
\end{tabular}
\end{minipage}
\end{centering}
\end{table*}

\begin{figure*}
\centering
\includegraphics[width = 10.0cm, angle=-90]{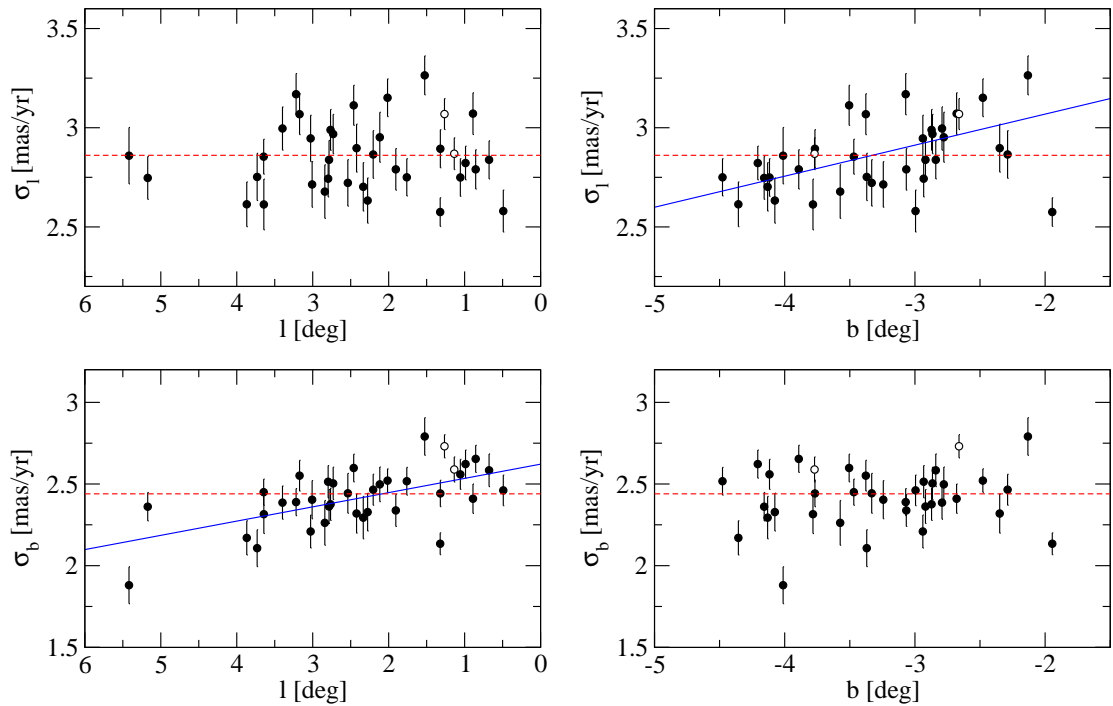}
\caption{Spatial dependence of proper motion dispersions $\sigmal$ and $\sigmab$ in Galactic coordinates
for our turn-off point dominated sample in the Galactic bulge (Table~\ref{tab:pm_results} and \S\,\ref{sec:results}).
The two open circles are for the Baade window and Sagittarius-I fields from Kuijken \& Rich (2002). The lines
show linear regressions (solid) and weighted means (dashed) of the data. For the top right panel,
the rightmost data point was not used in the fit.
}
\label{fig:dispersions}
\end{figure*}

\begin{figure*}
\centering
\includegraphics[width = 10.0cm, angle=-90]{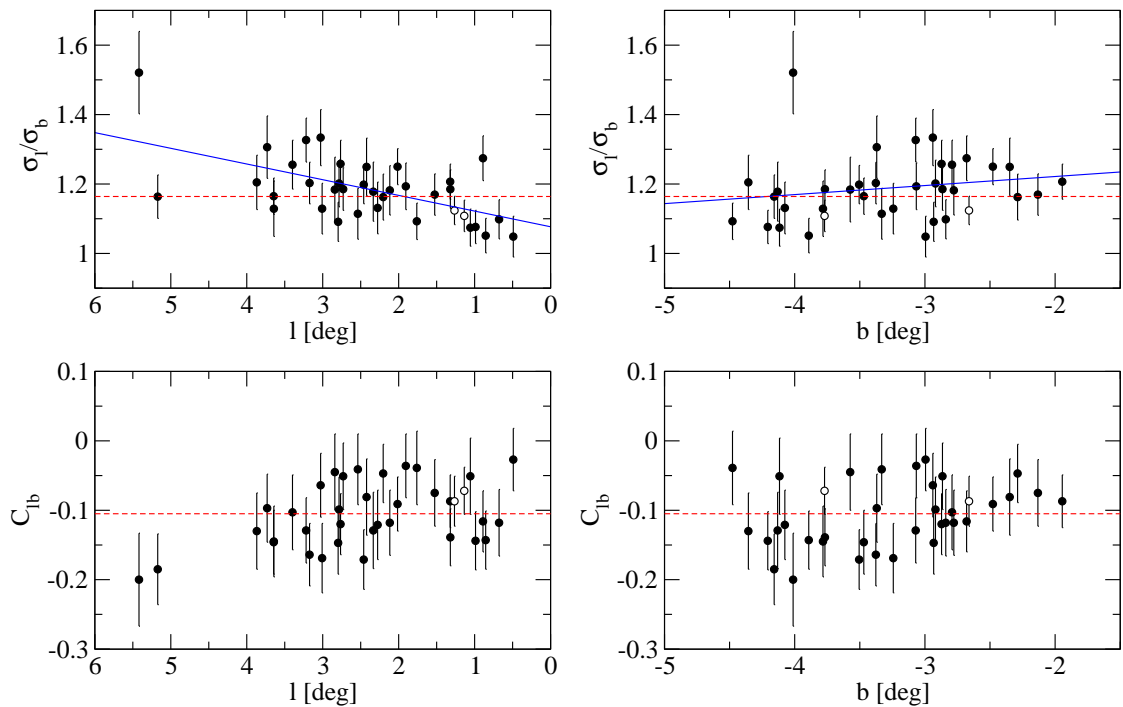}
\caption{Similar to Fig.~\ref{fig:dispersions} but for the anisotropy ratio $\sigmal/\sigmab$ and covariance term
$C_{\rm lb} \equiv \sigma_{\rm lb}/(\sigmal\sigmab)$.}
\label{fig:anisotropy}
\end{figure*}

\section{Results}
\label{sec:results}

The results are given in Table~\ref{tab:pm_results} and plotted in Figs.~\ref{fig:dispersions} and \ref{fig:anisotropy}.
Proper motions for individual stars are available online\footnote{\tt http://www.jb.man.ac.uk/\,\~\,simkoz/dispersions/}.
After presenting our measurements we check for consistency with two other published data sets (\S\,\ref{sec:cross_comp}).
A more detailed discussion and comparison to the results of \cite{KR02} follows in \S\,\ref{sec:discussion}.

\subsection{Proper motion dispersions}
\label{sec:dispersions}

The spatial dependence of proper motion dispersions $\sigmal, \sigmab$ is shown in Fig.~\ref{fig:dispersions}.
Recall that at a distance of 8\,kpc, a velocity of 100$\kms$ implies a proper motion of 2.64\,$\masyr$. The most visible trends
are in $\sigmal({\rm b})$ and $\sigmab({\rm l})$ that tend to increase closer to the Galactic plane and the Galactic centre.
Both gradients are weak, but clearly present. From a simple straight line fit we find: 

\begin{equation}
\sigmal = 0.16\pm0.04 \times{\rm b} + 3.38\pm0.13,
\label{eq:sigl_b}
\end{equation} 
\begin{equation}
\sigmab = -0.09\pm0.02 \times{\rm l} + 2.62\pm0.06 .
\label{eq:sigb_l}
\end{equation} 
Consequently, $\sigmab$ increases from 2.1 to 2.6\,$\masyr$,
or by about 20\%, as the longitude $\rm l$ varies from $5.5$ to $0.5$ deg. Similarly, $\sigmab$ changes from 2.6 to 3.2\,$\masyr$
between $\rm b = -4.5$ and $-2$ deg. It is intriguing that the last data point around $b \approx -2$ deg has 
the lowest dispersion ($\sigmal$) measured for all fields (see the top right panel in Fig.~\ref{fig:dispersions}),
but the value is still marginally consistent with the observed scatter. There is no other indication of the intrinsic variations
on the field-to-field scale. The distributions of $\sigmal({\rm l})$ and $\sigmab({\rm b})$ are flat within the scatter
from random errors and projection effects in the presence of gradients.

\subsection{Anisotropy and covariance}
\label{sec:anisotropy}

In Fig.~\ref{fig:anisotropy} we plot the ratio and the correlation coefficent (covariance of the velocity field) of $\sigmal$ and
$\sigmab$ as a function of location in the bulge.
There is a significant level of anisotropy, i.e. $\sigmal/\sigmab > 1$, throughout the covered area. Moreover, the velocity distribution
shows a tendency to become more isotropic for lines of sight approaching the Galactic centre at a roughly fixed latitude $\rm b$. This
is a reflection of the increase in $\sigmab$ with approximately constant $\sigmal$ (\S\,\ref{sec:dispersions}). The trend of more anisotropy
toward the Galactic plane is also driven primarily by one of the dispersions ($\sigmal$), but it is more difficult to see. Part of the
reason for this is the narrow range of $\rm b$ covered by the data. The formal fits give: 

\begin{equation}
\sigmal/\sigmab = 0.05\pm0.01 \times{\rm l} + 1.08\pm0.03,
\label{eq:siglb_l}
\end{equation} 
\begin{equation}
\sigmal/\sigmab = 0.03\pm0.03 \times{\rm b} + 1.27\pm0.08.
\label{eq:siglb_b}
\end{equation}
The estimates of the covariance term from Table~\ref{tab:pm_results} (plotted in Fig.~\ref{fig:anisotropy}) are all negative and scatter uniformly
in the range $-0.20 < C_{\rm lb} < -0.02$. This indicates that in our Galactic bulge fields the stellar motions in directions
parallel and perpendicular to the plane are significantly anti-correlated. An observational bias that would account for this anti-correlation
has to operate in a similar way over a large range of instrumental
settings. For example, a serious concern is a presence of preferred
telescope orientations. Indeed, for about half of our fields the relative roll angle between the two compared observations falls in a narrow
range of 25 deg. The other half, however, is spread over all possible orientations and still shows about the same covariance. The skewness
of the ACS focal plane cannot be the cause of the observed correlation, because the measurements in both \cite{KR02} fields use only WFPC2/PC
data and yet they perfectly agree with the rest of the $C_{\rm lb}$ values. 
We also investigated several other possibilities, we found no explanation for this result
other than a true correlation between $\mul$ and $\mub$. Taking a field with a relatively low S/N ratio in our data and assuming perfectly correlated
errors in $\mul$ and $\mub$, the expected covariance is only $C_{\rm lb} \sim 0.02$. 
There is a slight hint in Fig.~\ref{fig:anisotropy} that
$C_{\rm lb}$ may vary with the longitude, although this impression seems to rely on the two points farthest from the bulge minor axis (l$\simeq$5.3 
deg in the left panel of Fig.~\ref{fig:anisotropy}) that fall below the rest of the data.

\subsection{Comparisons with previous work}
\label{sec:cross_comp}

\subsubsection{OGLE-II proper motion catalogue}
\label{sec:OGLE2}

\begin{figure*}
\centering
\includegraphics[width = 8.0cm, angle=-90]{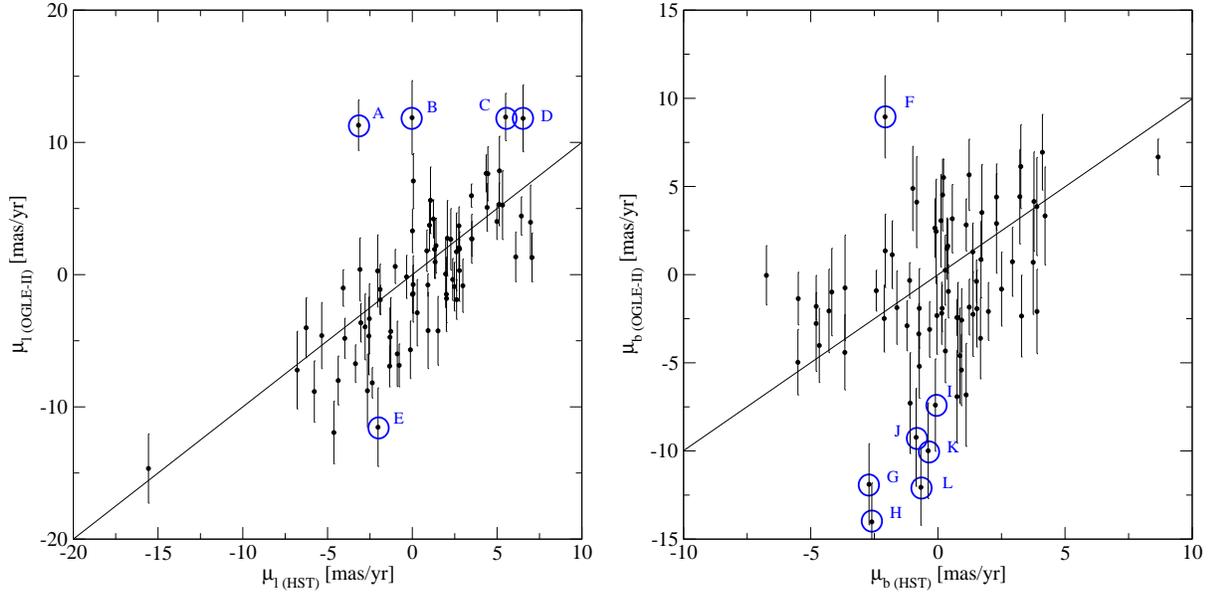}
\caption{Comparison between our HST measurements and the ground based OGLE-II data for bright stars 
from the catalogue of Sumi et al. (2004). There are 77 stars covered by our observations that have
catalogue errors 3 $\masyr$ or better in Sumi et al. Significant discrepancies (marked by alphabets) are caused by blending
(c.f Fig.~\ref{fig:stamps}). The solid lines indicate the two measurements (ground-based and from the HST) are equal.
}
\label{fig:sumi}
\end{figure*}

\cite{Sum04} used large number statistics of the OGLE-II database (\citealt{Uda97})
to derive relative proper motions of $\ga$5$\times10^6$ stars in the Galactic bulge region
from hundreds of observations covering a 4-year baseline. The OGLE-II catalogue
is a valuable resource for kinematic studies of bright stars like red clump giants
that are relatively free of the source confusion effects. However, at the
1.3$\arcsec$ FWHM seeing of the ground based OGLE-II images, a random red clump giant star
still has a few per cent probability of being blended with another unresolved star.
It is instructive to cross-validate the results of \cite{Sum04} and our high-resolution
HST measurements against each other.

\begin{figure}
\centering
\includegraphics[width = 8.3cm, angle=0]{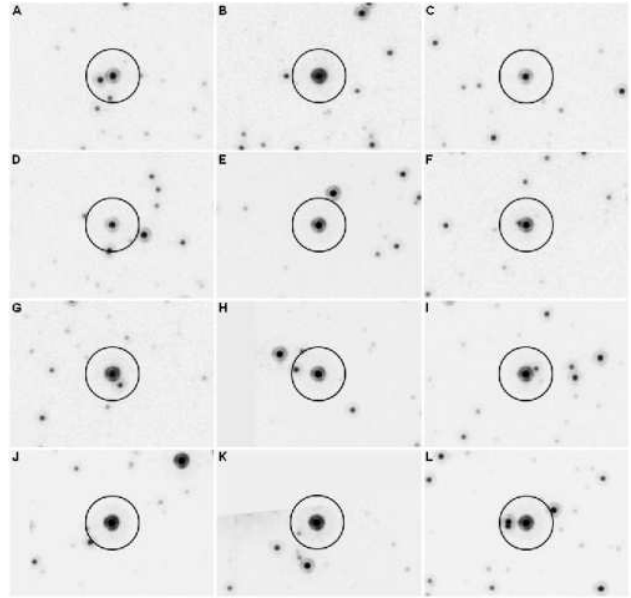}
\caption{Cutout HST images for the outliers marked by alphabets in Fig.~\ref{fig:sumi}. All significant outliers
in Fig.~\ref{fig:sumi} can be linked to source confusion and flux blending. The dark circles have a diameter of 1 arc second.
}
\label{fig:stamps}
\end{figure}

Out of 35 program fields in Table~\ref{tab:data}, 15 are covered by the OGLE-II proper motion
catalogue. In our HST sample we found 77 stars for which the OGLE-II catalogue proper
motion error is $\le 3\masyr$. The two data sets were compared star by star
after adjusting for an arbitrary zero point of the proper motion scale. The results
are plotted in Fig.~\ref{fig:sumi} and show a good overall agreement of our measurements
with \cite{Sum04}. All significant outliers were labeled and checked for blending.
Fig.~\ref{fig:stamps} demonstrates that virtually all these substantial discrepancies
are linked to the presence of an unresolved companion within $\sim$1$\arcsec$ of the primary object.

\subsubsection{Kuijken \& Rich (2002)}
\label{sec:KR2002}

Our approach to measuring the positions and proper motions of stars (\S\,\ref{sec:psf}) is somewhat simpler than the method
used by \cite{KR02}. The latter study utilized the images from the WFPC2/WF chips and had to accommodate a strong under-sampling
of the PSF. In contrast, our use of the WFPC2 data was limited to the critically sampled images from the PC detector.
The second epoch ACS/HRC images have four times the PSF sampling of the WF images, so we could take advantage of
the conventional PSF fitting techniques.

Regardless, in order to eliminate the possibility of a hidden error we re-analysed the PC data in both fields studied by \cite{KR02}
using our tools. Table~3 shows the results of this comparison. The agreement between the two sets of measurements
is remarkably close despite significant differences in the sample size and the adopted selection criteria. This also confirms
that our results are not significantly affected by several subtle instrumental effects that can potentially influence
astrometric work with the HST images (e.g. \citealt{KR02} and references therein).

\begin{table*}
\begin{centering}
\begin{minipage}{15cm}
\caption{Proper motion dispersions from Kuijken \& Rich (2002) compared with the results of our reanalysis of the same data}
\begin{tabular}{lcccccccccc}
\hline
&   &   & \makebox[0.5cm]{} & \multicolumn{3}{c}{This work} & \makebox[0.5cm]{} & \multicolumn{3}{c}{\cite{KR02}} \\*[0.1cm]
\cline{5-7} \cline{9-11}\\*[-0.2cm]
Field & l & b & & $\sigmal$ & $\sigmab$ & $N_{\rm stars}$ & & $\sigmal$ & $\sigmab$ & $N_{\rm stars}$ \\ 
\hline
BW      & 1.14 & $-$3.77 & & 2.87 $\pm$ 0.08 & 2.59 $\pm$ 0.08 & 694 & & 2.91 $\pm$ 0.06 & 2.51 $\pm$ 0.05 & 1076\\
Sgr-I   & 1.27 & $-$2.66 & & 3.07 $\pm$ 0.08 & 2.73 $\pm$ 0.07 & 752 & & 3.10 $\pm$ 0.06 & 2.73 $\pm$ 0.05 & 1388\\
\hline
\end{tabular}
\end{minipage}
\end{centering}
\label{tab:KRcomp}
\end{table*}

\section{Discussion}
\label{sec:discussion}

\begin{figure*}
\centering
\includegraphics[width = 8.0cm, angle=-90]{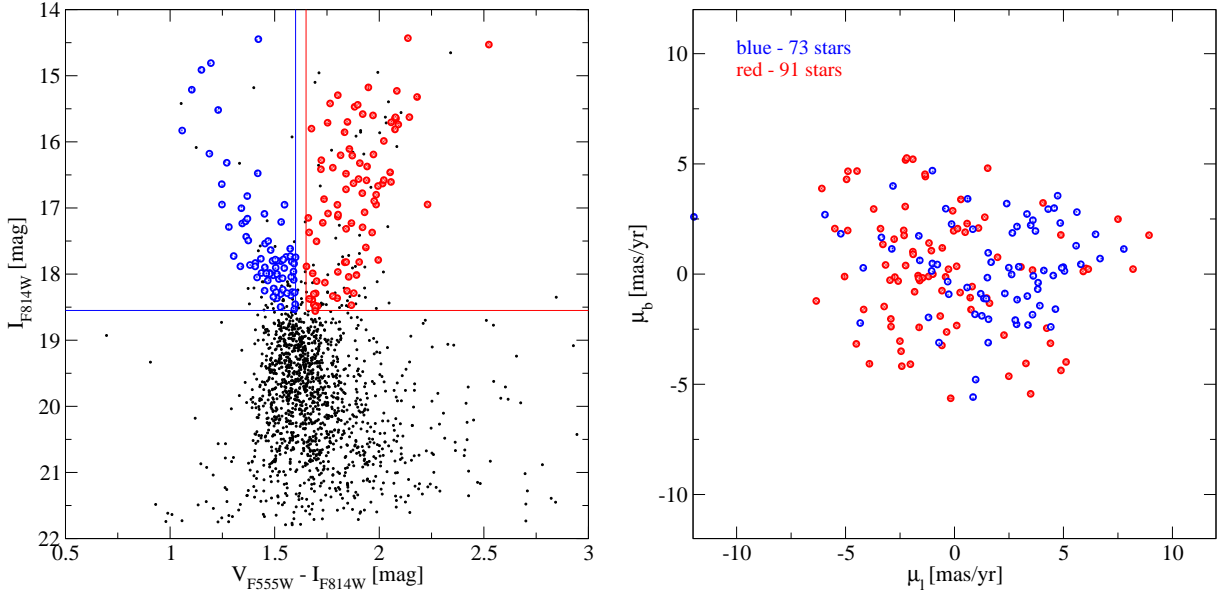}
\caption{Color-magnitude diagram (CMD; left) and relative proper motions (right) for stars in three nearly coincident stellar fields
from Table~\ref{tab:pm_results}: 97-BLG-18, 104-C and 104-D. The red and blue stars above the turn-off point show the kinematics
characteristic of the bulge and disk populations, respectively.
}
\label{fig:cmd_pm_dist}
\end{figure*}

\begin{figure*}
\centering
\includegraphics[width = 8.0cm, angle=0]{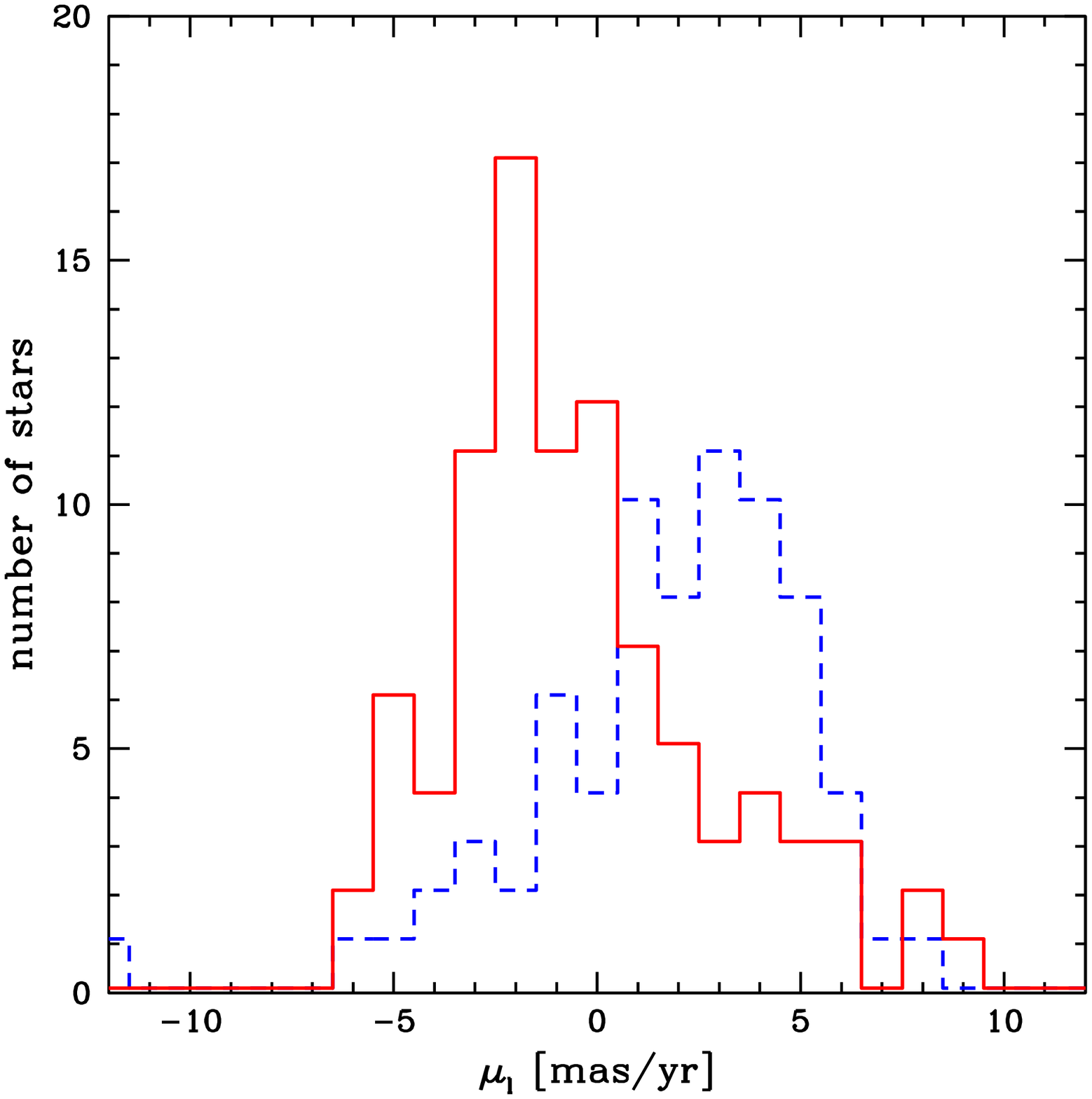}\includegraphics[width = 8.0cm, angle=0]{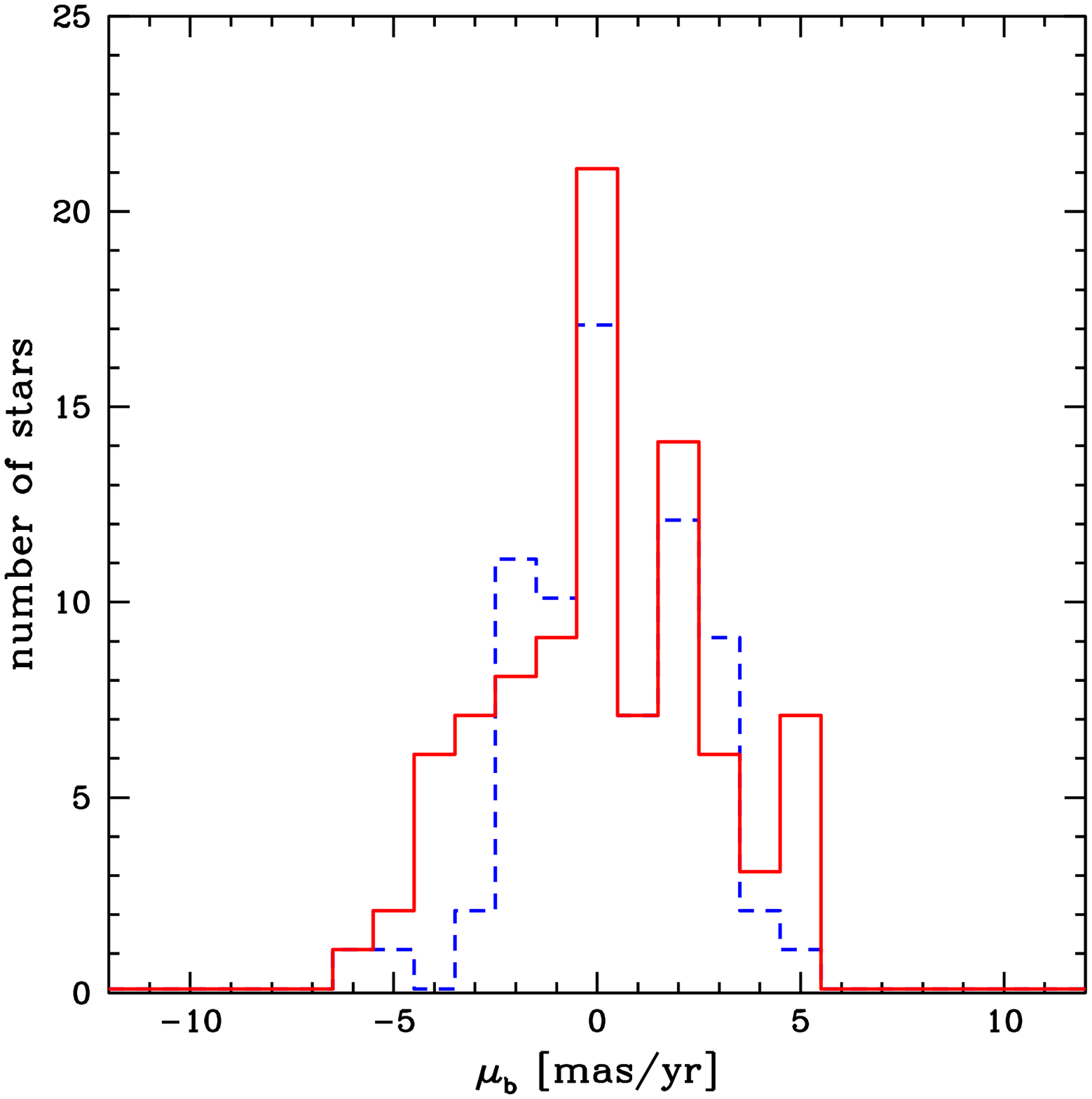}
\caption{Histograms of relative proper motions of the red (solid line) and blue (dashed line) samples from Fig.~\ref{fig:cmd_pm_dist}
The blue disk stars ``rotate in front'' of the Galactic bulge parallel to the plane.}
\label{fig:pm_dist}
\end{figure*}

\begin{figure*}
\centering
\includegraphics[width = 8cm, angle=-90]{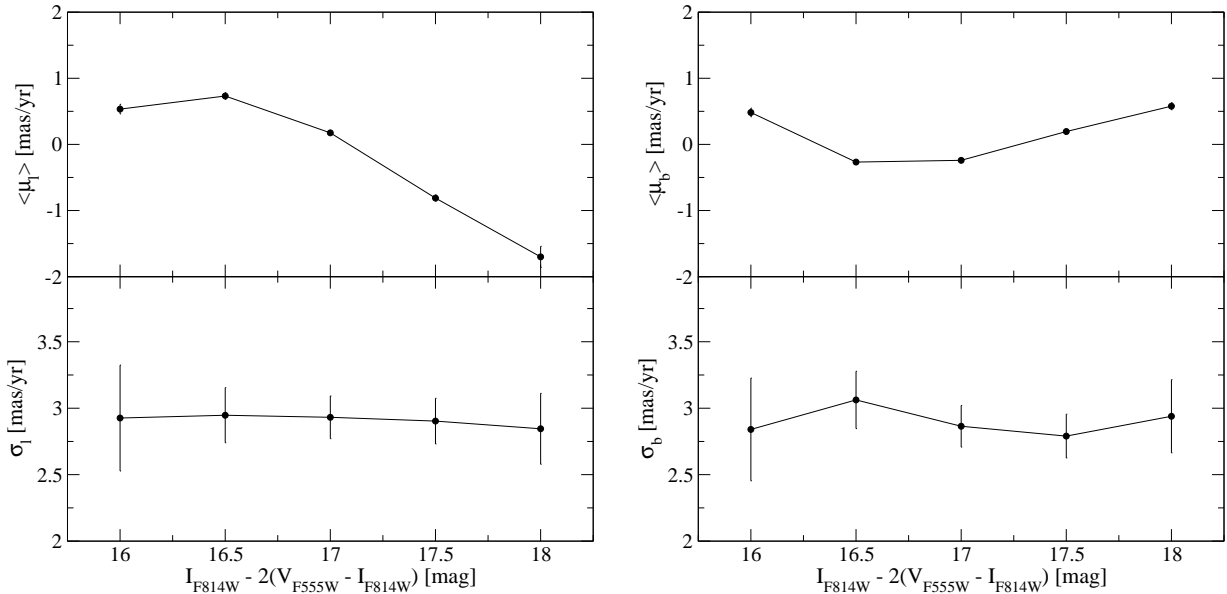}
\caption{Average relative proper motions and dispersions of stars in the
Baade Window in bins of $M^* \equiv I_{\rm F814W}-2\times(V_{\rm
F555W}-I_{\rm F814W})$,
an approximate distance indicator.}
\label{fig:KR2}
\end{figure*}

\subsection{Distance and population trends}
\label{sec:trends}

The study of \cite{KR02} focused on cleaning the Galactic bulge population and removing
the contamination by the bluer disk stars. Above the bulge turn-off point, the stellar colors alone
are sufficient to separate the blue disk main sequence from the red giants, subgiants and clump
giants. The size of our fields is generally too small to provide useful statistics of bright
stars above turn-off point, and good color information is only available for about 1/3 of the lines of sight.
However, three of the fields in Table~\ref{tab:data} with useful colors (97-BLG-18, 104-C and 104-D)
are close to each other and were combined in order to look for a kinematic distinction
between the Galactic disk and bulge populations. Figs~\ref{fig:cmd_pm_dist} and \ref{fig:pm_dist}
show that in the frame of reference of a mean star (of any color), the longitude proper motions
of the blue disk stars are biased toward positive values, while the red bulge stars tend to have
more negative $\mul$. The blue and red samples were selected, correspondingly,
using conditions $(V-I) < 1.65$ and $(V-I) > 1.7$. This effect was previously observed
by \cite{Sum05} and by \cite{KR02} in their two fields with multi-epoch WFPC2 data and there is little doubt that
it is due to the blue disk stars ``rotating in front'' of the red bulge stars.

\cite{KR02} also devised an approximate distance measure:
\begin{equation}
M^* = I_{\rm F814W}-2(V_{\rm F555W}-I_{\rm F814W})
\end{equation}

\noindent
chosen to remove the slope of the main sequence in the color-magnitude diagram. In Fig.~\ref{fig:KR2}
we present the average proper motions and their dispersions for the Baade window in bins of $M^*$.
As expected, with an increasing depth along the line of sight, the kinematic signature gradually changes from
that characteristic of the disk stars, to the one typical for the bulge. In the \cite{KR02} data this trend
continues to very faint stars that are likely on the far side of the bulge, and if so, it constitutes
a ``rotation curve'' of the bulge. The colors for our fields are generally of lower S/N ratio or nonexistent,
and do not allow to see this much detail.

\subsection{Stellar velocity ellipsoid of the Galactic bar}
\label{sec:covar}

A detailed modeling of the measurements in Table~\ref{tab:pm_results} is beyond the scope of this paper.
Here we only comment on possible directions and new possibilities.

\cite{ZRB96} interpreted the bulge anisotropy in terms of the rotation support of the Galactic bulge and related
the ratio $\sigmal/\sigmab$ directly to the level of flattening of the light density distribution.
They also concluded that the value $\sigmal/\sigmab = 1.10$--1.15 observed in Baade window
$({\rm l}, {\rm b}) \sim (1^\circ, -4^\circ)$, with which our measurements are consistent, can be explained
by rigid rotation. The presence of any disk stars, however, will also contribute rotational broadening
to $\sigmal$. Since in the vicinity of our fields the disk fraction increases closer to the plane,
it follows that the measured gradient from equation~(\ref{eq:sigl_b})
could be due to the disk contamination. The changes of skewness in the $\mul$ distribution tend
to support this (see Fig.~\ref{fig:pm_dist}).
Another possibility is that the rotation rate of the bulge actually
increases at lower $|b|$, as found by \cite{Izu95} from the radial velocities of 124 SiO masers in the Galactic bulge.
It has been observed that for giants in Baade window the metal-poor stars display more spread
in the vertical motion and less anisotropy when compared to metal-rich samples (\citealt{ZSR94,ZRB96}).
Both of these metallicity dependencies are quite steep, so it is likely that the gradient
from equation~(\ref{eq:sigb_l}) is related to a changing mix of populations
with more metal-poor stars closer to the Galactic bulge minor axis.

We are not aware of any previous
detections of the cross terms in the Galactic bulge velocity field except the report by \cite{ZSR94}
of a significant vertex deviation between the radial and longitudinal motions from $C_{\rm rl}$.
That result is based on a photographic sample of $\sim$200 K and M giants from \cite{Spa92}.
We note that the latter sample actually shows a hint of a slightly negative covariance between $\mul$ and $\mub$
(c.f. Fig.~1 of \citealt{ZSR94}). The superb resolution of HST enabled very significant detections of the
$C_{\rm lb}$ cross term in many fields. The non-diagonal elements of the velocity tensor are crucial
to determining the dominant orbit families, the importance of streaming motions and the need for
the intrinsic anisotropy versus solid body rotation in the Galactic bulge (\citealt{ZSR94,ZRB96,Haf00}).
\cite{Haf00} published detailed calculations of $C_{\rm lb}$ for several lines of sight at positive longitudes
including Baade window $(1^\circ, -4^\circ, C_{\rm lb} = 0.04)$, and two other:
$(8.4^\circ, -6^\circ, C_{\rm lb} = 0.15)$ and $(1.21^\circ, -1.67^\circ, C_{\rm lb} = 0.04)$.
Taken at face value these predictions are roughly of the same magnitude as the results from \S\,\ref{sec:results},
but have the opposite sign. For a proper comparison with dynamical models like the ones in \cite{Haf00}
and \cite{Bis04} we need to wait until the calculations are folded with the appropriate selection
functions, since our measurements are based on substantially deeper data than most of the previous samples.

\section{Summary and Conclusions}
\label{sec:conclusions}

The main results of our proper motion mini-survey are: (1) high quality proper motion measurements
for hundreds of stars in 35 lines of sight across the Galactic bar, (2) establishing the presence of spatial
gradients in dispersions $\sigmal, \sigmab$ and the amount of anisotropy $\sigmal/\sigmab$, and (3) the first
reliable detection of the covariance term $C_{\rm lb}$ of the transverse velocity tensor. We cross-validated
our measurements with the ground based OGLE-II data of \cite{Sum04} and a benchmark study of \cite{KR02}.
The observed slow rise of $\sigmal$ toward the Galactic plane is likely due to the increasing disk contamination
and/or a possible gradient in the bulge rotation speed. The increase in $\sigmab$ toward the minor axis of the bulge
is accompanied by the decreasing ratio $\sigmal/\sigmab$ and possibly results from increasing
fraction of metal-poor stars. Another possibility is that $\sigmab$ increases due to the larger surface density
of stars at low $l$ (closer to the Galactic centre).
We clearly detect the covariance term $C_{\rm lb} \sim -0.10$ that implies
a significant tilt of the Galactic bulge velocity ellipsoid with respect to the Galactic plane.
Using the same procedures as in \cite{BM98}, we find the tilt is roughly equal to $-24^\circ$.

The data presented in this paper provide qualitatively new constraints on dynamical models of the inner
Galaxy and dramatically improved number statistics. Furthermore, it may be possible in the near future to augment
our proper motion samples with the distance and metallicity estimates. As shown by \cite{KR02}, deep
color-magnitude diagrams can supply sufficiently accurate distance information to effectively
isolate the bulge population. In order to maximize the discriminating
power of model comparisons the focus should be on extending the coverage to negative longitudes and locations
further from the Galactic centre.

\section*{Acknowledgments}
We thank Profs. Bohdan Paczy\'nski, Ian Browne and Drs. Vasily Belokurov, Wyn Evans and Nicholas Rattenbury for helpful comments.

Support to P.W. for proposal SNAP-10198 was provided by NASA through a grant from the Space Telescope Science Institute,
which is operated by the Association of Universities for Research in Astronomy, Inc., under NASA contract NAS5-26555.

This work was partly supported by the European Community's Sixth Framework Marie Curie Research Training Network Programme,
Contract No. MRTN-CT-2004-505183 ``ANGLES'', in particular S.K. through a studentship, 
L.W. through a PDRA, and S.M. through travel support.

M.C.S. acknowledges financial support from the Netherlands Organisation for Scientific Research (NWO).

\bibliographystyle{mn2e}

\end{document}